\documentclass[aps,prl,reprint,superscriptaddress,showpacs]{revtex4-1}

\usepackage[utf8]{inputenc}

\usepackage{amsbsy}
\usepackage{amsfonts}
\usepackage{amssymb}
\usepackage{amsmath}
\usepackage{color}
\usepackage{graphicx}
\usepackage{float}
\usepackage[caption = false]{subfig}

\begin{document}


\title{A Nonequilibrium Quantum Phase Transition in Strongly Coupled Spin Chains}

\author{Eduardo Mascarenhas}
\affiliation{Institute of Physics, Ecole Polytechnique F\'{e}d\'{e}rale de Lausanne (EPFL), CH-1015 Lausanne, Switzerland}

\author{Giacomo Giudice}
\affiliation{Institute of Physics, Ecole Polytechnique F\'{e}d\'{e}rale de Lausanne (EPFL), CH-1015 Lausanne, Switzerland}

\author{Vincenzo Savona}
\affiliation{Institute of Physics, Ecole Polytechnique F\'{e}d\'{e}rale de Lausanne (EPFL), CH-1015 Lausanne, Switzerland}

\begin{abstract}
We study spin transport in a boundary driven XXZ spin chain. Driving at the chain boundaries is modeled by two additional spin chains prepared in oppositely polarized states. Emergent behavior, both in the transient dynamics and in the long-time quasi-steady state, is demonstrated.
Time-dependent matrix-product-state simulations of the system-bath state show ballistic spin transport below the Heisenberg isotropic point. Indications of exponentially vanishing transport are found above the Heisenberg point for low energy initial states while the current decays asymptotically as a power law for high energy states. Precisely at the critical point, non-ballistic transport is observed. Finally, it is found that the sensitivity of the quasi-stationary state on the initial state of the chain is a good witness of the different transport regimes.
\end{abstract}

\maketitle

Nonequilibrium phase transitions have emerged as critical phenomena that may underpin novel forms of universality, departing significantly from transitions that are driven by thermal or quantum fluctuations~\cite{Sachdev}. A prototypical setting for these critical phenomena is that of transport in one-dimensional systems, and in particular many-body spin chains such as the spin-1/2 XXZ chain. This model system is known to provide an accurate description of real materials~\cite{ExNatPhys,XXZmat,XXZmat2,HessWTVs} and can also be quantum-simulated in superconducting circuits~\cite{XXZSupCond,XXZSupCond2,XXZSupCond3}. With the outstanding developments and control in trapped ions and ultracold atoms in optical lattices~\cite{Lattice,Lattice2,Lattice3,Lattice4,Bloch,Monroe,PhysRevLett.113.147205,mimimi,mimimi2,mimimi3,mimimi4,mimimi5,FTR,FTR2} and molecules~\cite{MulecaNat,MulecaPRL,MulecaPRA}, similar models can also be simulated in such architectures.

Transport through the XXZ spin chain has been investigated in several studies, both from the perspective of linear response (i.e., the Kubo formalism)~\cite{PhysRevB.55.11029,PhysRevLett.106.217206,PhysRevLett.111.057203,PhysRevB.80.184402,PhysRevLett.107.250602} and from the point of view of quench-induced dynamics~\cite{PhysRevE.59.4912,PhysRevE.57.5184,PhysRevE.81.061134,PhysRevE.71.036102}. These works have highlighted the possibility of both ballistic and diffusive transport of the spin current and characterized the transition between the different regimes. 

More recently, the study of nonequilibrium critical phenomena has been extended to open quantum systems, where the nonequilibrium character is induced by coupling the system to several external reservoirs~\cite{PhysRevLett.116.070407}. Theoretical investigation of open quantum systems is ultimately motivated by the inherently open nature of several modern experimental platforms~\cite{Revhart}, which are typically subject to external drive, dissipation and dephasing. One of the main goals of these studies is to investigate the possibility of critical phenomena that differ substantially from those occurring in closed, Hamiltonian systems, possibly leading to the emergence of new universality classes~\cite{PhysRevLett.116.070407}.

In the context of open quantum systems, the problem of spin transport has been addressed by assuming {\em boundary driven} spin chains, i.e. chains in which only the first and last spins are coupled to external environments which induce the necessary bias to trigger transport. Boundary driven many-body systems are rapidly emerging as a new paradigm for nonequilibrium critical phenomena, and their investigation may shed light on the fundamental mechanisms underlying many-body transport in more complex systems~\cite{Casati}. The boundary driven XXZ spin chain in particular has recently been the object of several studies, as it displays distinct nonequilibrium transport behaviour depending on the driving bias and on the coupling anisotropy~\cite{ZnidaricLowu,Prosen}.

Transport in the boundary driven XXZ chain has been studied under the coupling to Markovian spin baths, both in the linear response regime close to infinite temperature~\cite{ZnidaricLowu} and far from equilibrium in the presence of full bias~\cite{Prosen}. In the low bias (linear) regime, a transition from ballistic to diffusive transport has been shown at the Heisenberg isotropic point~\cite{ZnidaricLowu}. At full bias instead, the isotropic spin chain is characterized by a subdiffusive transport regime which separates the ballistic from an insulating regime, respectively below and above the Heisenberg point~\cite{Prosen}. Other theoretical investigations of transport through the boundary driven XXZ spin chain have studied the possible analogies with the closed-system counterpart~\cite{FTR3}, the onset of local thermalization~\cite{PhysRevE.91.042129}, the effects of additional bulk dephasing~\cite{PhysRevB.87.235130}, and the existence of a negative differential conductivity regime~\cite{PhysRevB.80.035110,Nega}. An insulating-type behavior at large bias was reported in Ref.~\cite{Nega}. In all these studies, an important assumption is that of weak, Markovian coupling of the system to the baths, which can be modeled using a quantum master equation formalism or an equivalent quantum trajectory approach.

\begin{figure}
{\includegraphics[width = 3.4in]{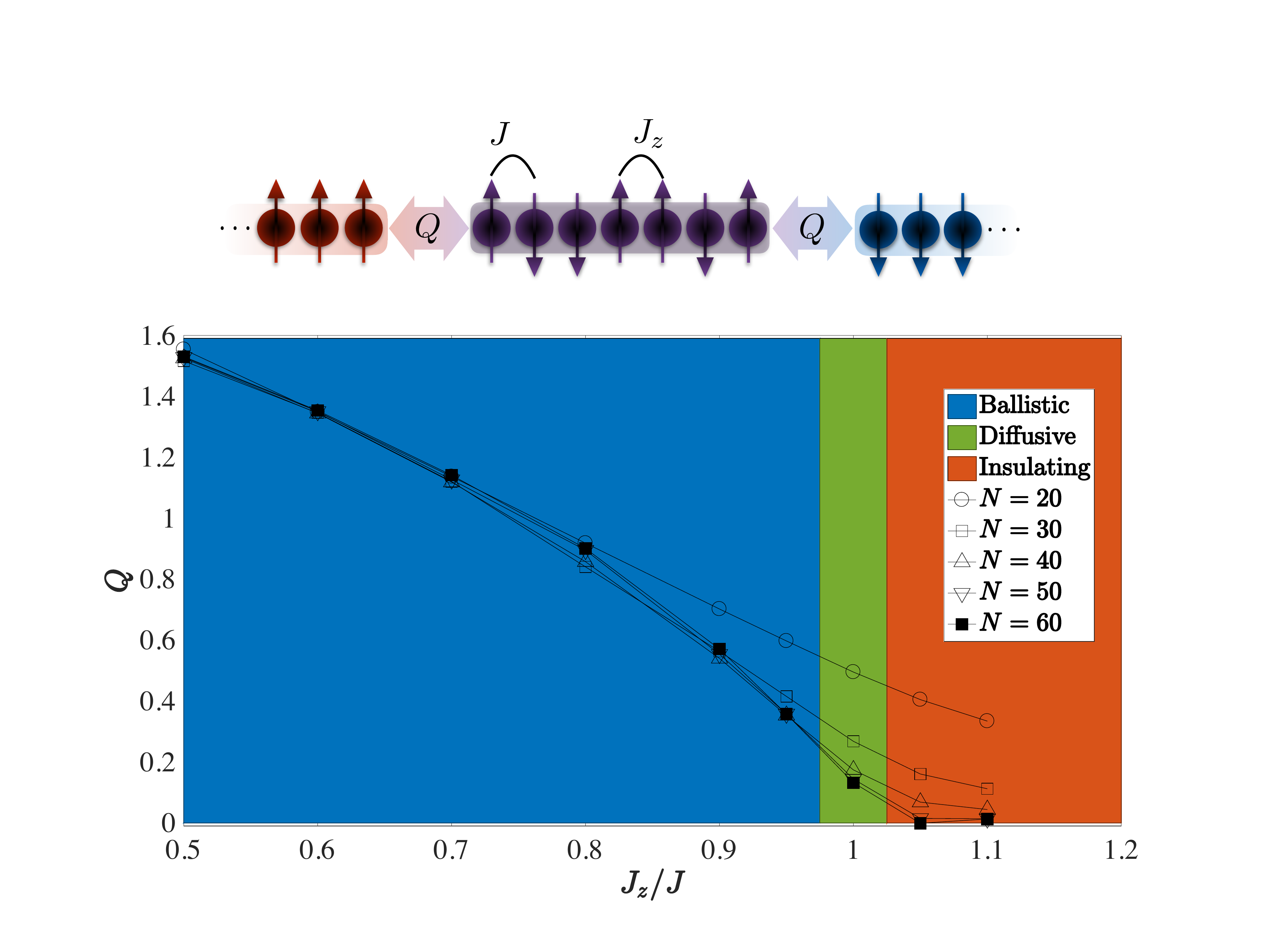}}\\ 
\caption{(Above) Drawing of the battery model. A system is coupled to two leads that are fully magnetized in opposite directions. The time evolution generates a current $Q(t)$ between the system and the battery. (Below) Sketch of the regime diagram inferred from the simulations. The superimposed data represent the quasi-steady-state current $Q(\tau_2)$, with $\tau_2=0.5N/J$, as computed for different system sizes as a function of the $J_z$ coupling.}
\label{ChainWBathsPhase}
\end{figure}

\begin{figure*}
{\includegraphics[width = 6.5in]{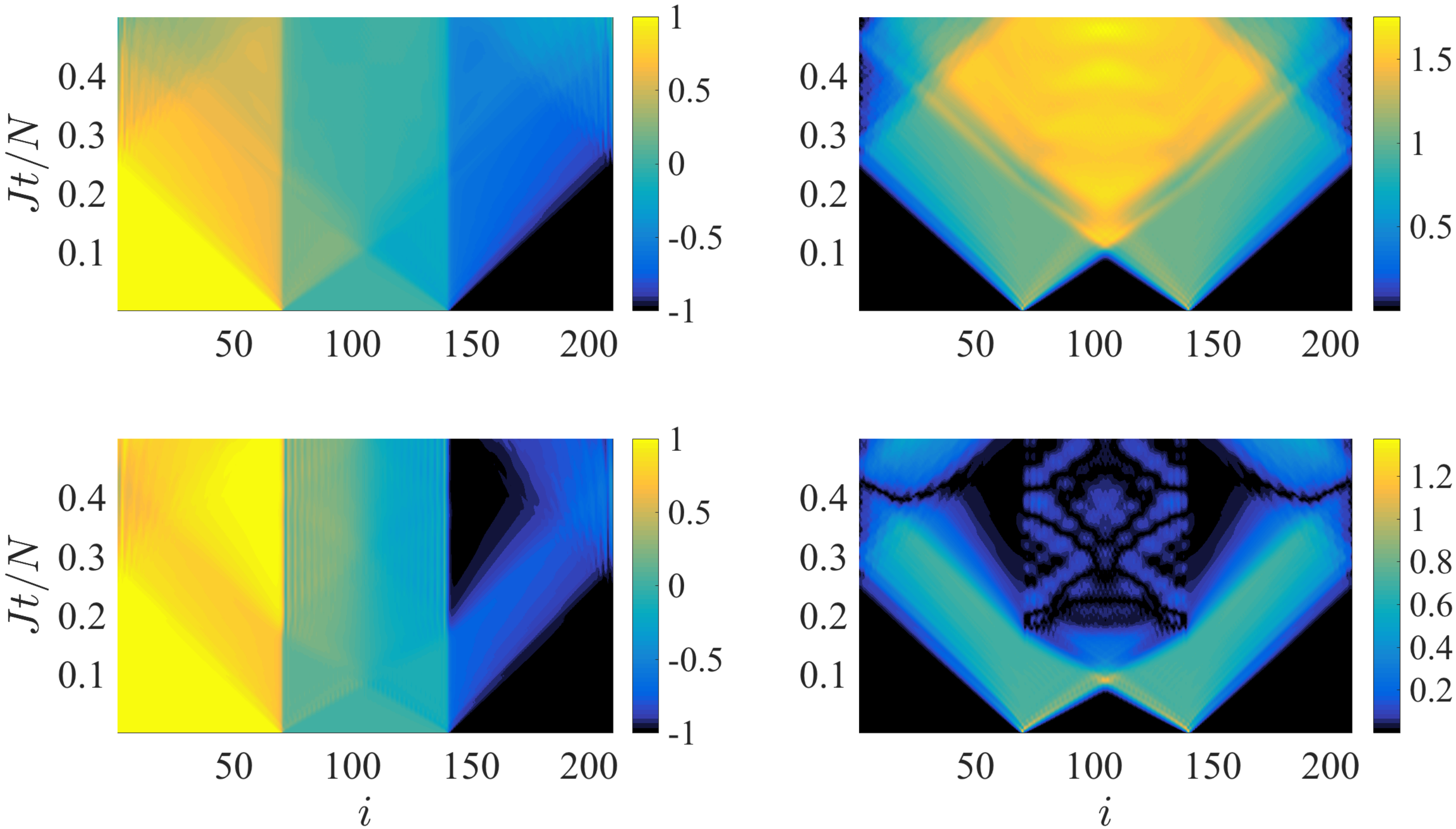}}\\ 
\caption{Space and time dependent color plot of the magnetization $\langle Z_i \rangle$ on the left panels and spin current $Q_i$ on the right panels for a system of $N=70$ sites (the total number of sites is $210$). Upper panels: $J_z=0.5J$. Lower panels: $J_z=1.1J$.}
\label{ChainWBathsN60Jz0p5Jz1p1}
\end{figure*}

A key question is whether dropping the weak coupling assumption and considering strong coupling between the chain and the leads may lead to significantly different transport properties and, possibly, to a new critical behavior. To address this question, a possible strategy consists in extending the phenomenological master equation approach to the non-Markovian case~\cite{OpenBreuer,InesRev}. Alternatively, several studies have adopted the more pragmatic approach of modeling the main chain and (finite size) baths as a unique Hamiltonian system, in which system and baths may be prepared in different initial states and the subsequent unitary evolution studied in the limits of long chain lengths and times. In this respect, the system of two spin chains coupled by a single junction and subject to local quenches has been thoroughly investigated. If the global Hamiltonian describing system and bath is homogeneous, then the system is solvable by Bethe Ansatz, and this case has recently been studied under generalized hydrodynamics assumption~\cite{DeLuca}. In this setting, a transition from ballistic to diffusive transport across the junction was recently demonstrated~\cite{SJZP}. Other studies have investigated light cone velocities~\cite{Lauchli}, entanglement spreading~\cite{geometric}, nonequilibrium steady states~\cite{NonequilibriumS}, the time evolution of two chains at different temperatures~\cite{LucaMaluca,TempFermi,Biella,Karrasch,Collura}, the comparison between Landauer, Kubo, and microcanonical descriptions~\cite{Zwolak}, and the emergent hydrodynamics~\cite{Hydra1,Hydra2}. Furthermore, in line with the results obtained for the Markov case in the low bias regime, ballistic to diffusive transport was found in the limit of infinite temperatures~\cite{Meisner}. 

Here, we specifically address the task of characterizing critical phenomena in the transport through a XXZ spin chain strongly coupled to two non-Markovian baths at its boundaries. To this purpose, we extend the single-junction approach and model the system and the two baths as three coupled spin chains~\cite{CU}. We explicitly model the full unitary evolution of the system-bath dynamics, in order to circumvent the phenomenological master equation approach. In this setting, the baths operate as a magnetization battery, driving the spin transport through the system. Modeling the baths as spin chains results in a finite energy bandwidth for their eigenspectrum. We therefore expect this model to display significant non-Markovian effects in the system-bath intraction, as a Markovian dynamics would strictly require an environment with a ``white" spectrum~\cite{InesBath,Prior}.

More specifically, we prepare one lead of the battery in the fully up-magnetized state, denoted by $|11...111\rangle$, and the other lead in the down-magnetized state, denoted by $|00...000\rangle$. 
The initial state of the whole system and baths is therefore $|\Psi_S\rangle=|11...111\rangle|S\rangle|00...000\rangle$, with $|S\rangle$ being the initial state chosen for the system. The system itself is prepared in a state with zero total magnetization, which in most of the following analysis is assumed to be $|S\rangle=|G\rangle$, the ground state of the Hamiltonian. This initial configuration $|\Psi_G\rangle$ induces a strongly nonlinear transport response, resembling the maximal bias regime in the case with open, Markovian boundary leads~\cite{Prosen}.
Lower magnetization differences between the leads of the battery eventually lead to the linear response regime at low bias~\cite{ZnidaricLowu}.

As opposed to the Markovian open-system setting~\cite{Prosen,ZnidaricLowu}, here the system-battery coupling does not in general map to a simple \textit{local} phenomenological master equation. Deriving a closed form master equation for strongly interacting systems requires, in principle, knowledge of the full eigen-decomposition of the system Hamiltonian~\cite{OpenBreuer} --- a hopeless task for a many-body problem. It was also shown recently that the local phenomenological approach is not able to capture complex transport phenomena such as edge currents~\cite{Rivas}. Finally, it was recently pointed out that, when in presence of multiple baths, the approach consisting in including each bath additively in the master equation is inherently inaccurate, as interference terms may arise, with significant effect on the ensuing non-equilibrium steady state~\cite{MarkMitch}.
Therefore, modeling the full Hamiltonian dynamics of the system and the baths is a convenient way to correctly address the strong system-bath coupling, the strongly many-body character of the system, and memory effects~\cite{IniState}. In this non-Markovian setting, an important question is to what extent the critical behavior depends on the choice of the initial state of the  system. We therefore repeat part of the analysis under the assumption that the system is initially prepared in the generalized GHZ state $|GHZ\rangle=(|11...111\rangle+|00...000\rangle)/\sqrt{2}$, which is orthogonal to the state $|G\rangle$ while also having zero total magnetization. With these the two initial states we have considered, we find that for given system parameters the long-time current behaves essentially in the same manner, thus characterizing the same transport regime. However, especially in the insulating regime, the initial state can have a strong influence on the actual microscopic state reached by the system in the long-time limit. We confirm this statement by studying the overlap between the time-evolution of different initial states.

\begin{figure}
{\includegraphics[width = 3.4in]{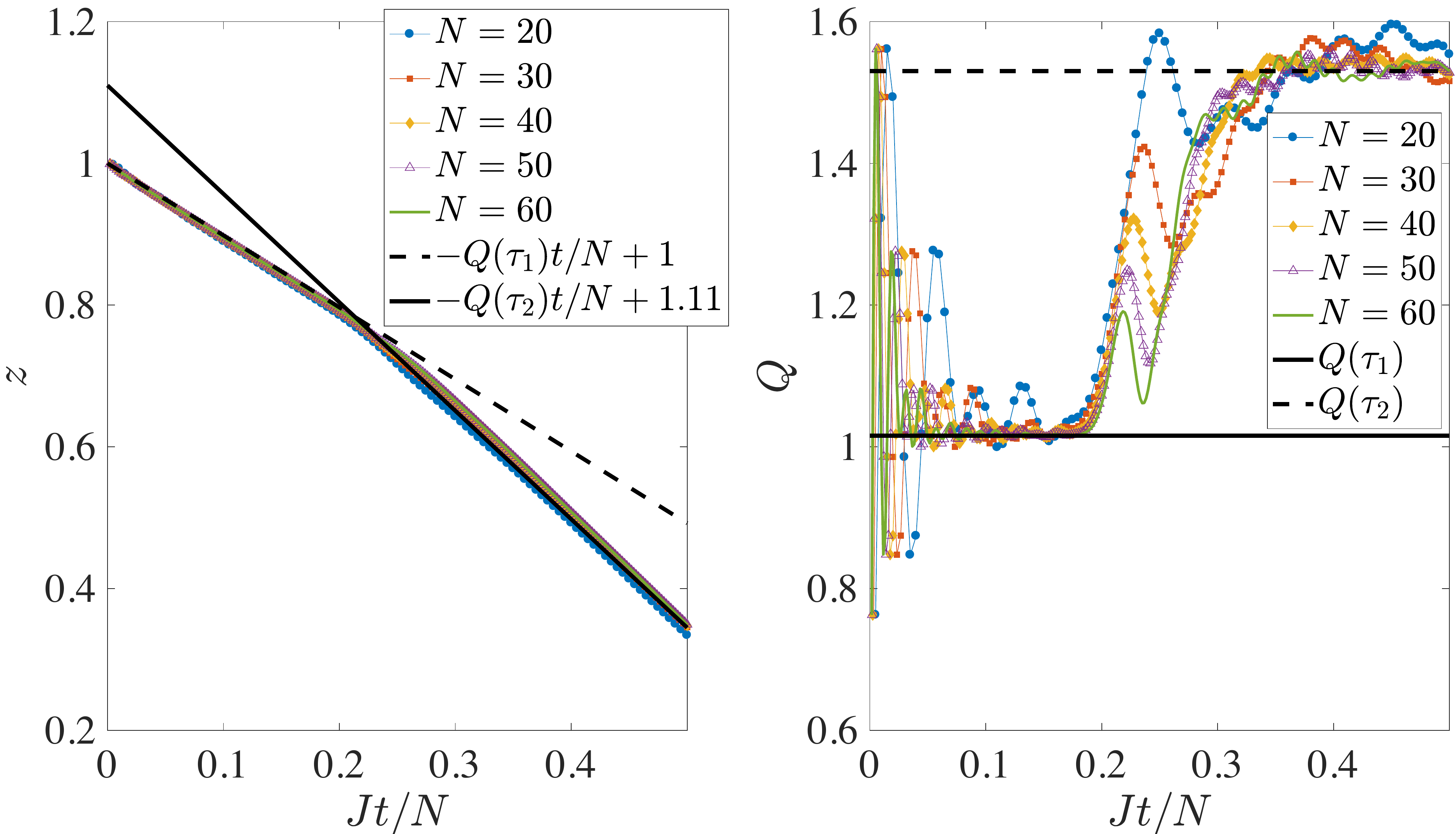}}\\ 
\caption{ (Left) The battery magnetization as a function of time, for different system sizes and (Right) the current at the battery-system junction, as computed for $J_z=0.5J$. The values $Q(\tau_i)$ were obtained assuming a system size $N=60$ with $\tau_1=0.1N/J$ and $\tau_2=0.5N/J$.}
\label{ChainWBathsJz0p5}
\end{figure}

The Hamiltonian of the XXZ spin chain, expressed in terms of the Pauli matrices $\{X,Y,Z\}$ is
\begin{equation}H_{\mathrm{XXZ}}=\sum_i^{N-1} J(X_iX_{i+1} +Y_iY_{i+1})+J_zZ_iZ_{i+1},\end{equation}
where $J$ is the spin hopping rate, $J_z$ is the spin repulsion and $N$ is the number of spins in the chain. 
We assume that the chains representing the batteries are ballistic conductors with $J_z=0$ and that the coupling to the system is also an XX (or hopping) Hamiltonian. The number of spins in the bath chains is $N_b\ge N$. This relation between system and bath sizes is chosen to avoid strong recurrences before the quasi-steady state is reached.
Since we have a non-homogeneous global Hamiltonian, Bethe Ansatz techniques are not currently available. On the other hand, the protocol outlined is a combination of two local quenches and DMRG has a consistent history of efficiently simulating local quenches~\cite{SJZP,Lauchli,geometric,NonequilibriumS,Biella,Meisner}.
Therefore, we simulate the model with time-dependent matrix product states techniques (t-MPS)~\cite{DMRGMPS,VMPS}. The dynamics is simulated adopting a symmetric fourth-order Trotter-Suzuki splitting with time step ${\rm d}t=0.1/J$
and with maximal bond dimension $D=500$ which keeps the accumulated errors under control. 
{We apply a variational compression to the MPS that minimizes the infidelity between the uncompressed and compressed states. The maximal infidelity we registered was of the order of $10^{-4}$, making our results reliable in a wide range of parameters.}
We perform finite-size scaling of the spin current, which enables the characterization of the three different transport regimes.

The spin current at each link of the chain $Q_i=2J\langle X_iY_{i+1}-Y_iX_{i+1}\rangle$ is defined through the continuity equation $\dot{\langle Z_i\rangle}=Q_{i-1}-Q_i$. 
The time-dependent current measured between the system and one lead of the battery $Q(\tau)$ displays two distinct transient regimes, extending respectively up to times $\tau_1$ and $\tau_2$, that will be discussed below. After these transients, a quasi-stationary current $Q(\tau_2)$ can be identified and associated to the quasi-steady state. The summary of our results is presented in the regime diagram shown in Fig.~\ref{ChainWBathsPhase}, where we plot the bare value of the currents $Q(\tau_2)$ for different system sizes. Ballistic transport is found for $J_z<J$ and is characterized by currents that are finite even in the thermodynamic limit, where they asymptotically approach a constant value as a function of the system size. Non-ballistic transport, including anomalous diffusion, is found at the Heisenberg point, and is characterized by an algebraic scaling $Q=A N^{-\alpha}$. The insulating regime found for $J_z>J$ displays instead an exponential scaling $Q=B e^{-\beta N}$.

\begin{figure}
{\includegraphics[width = 3.4in]{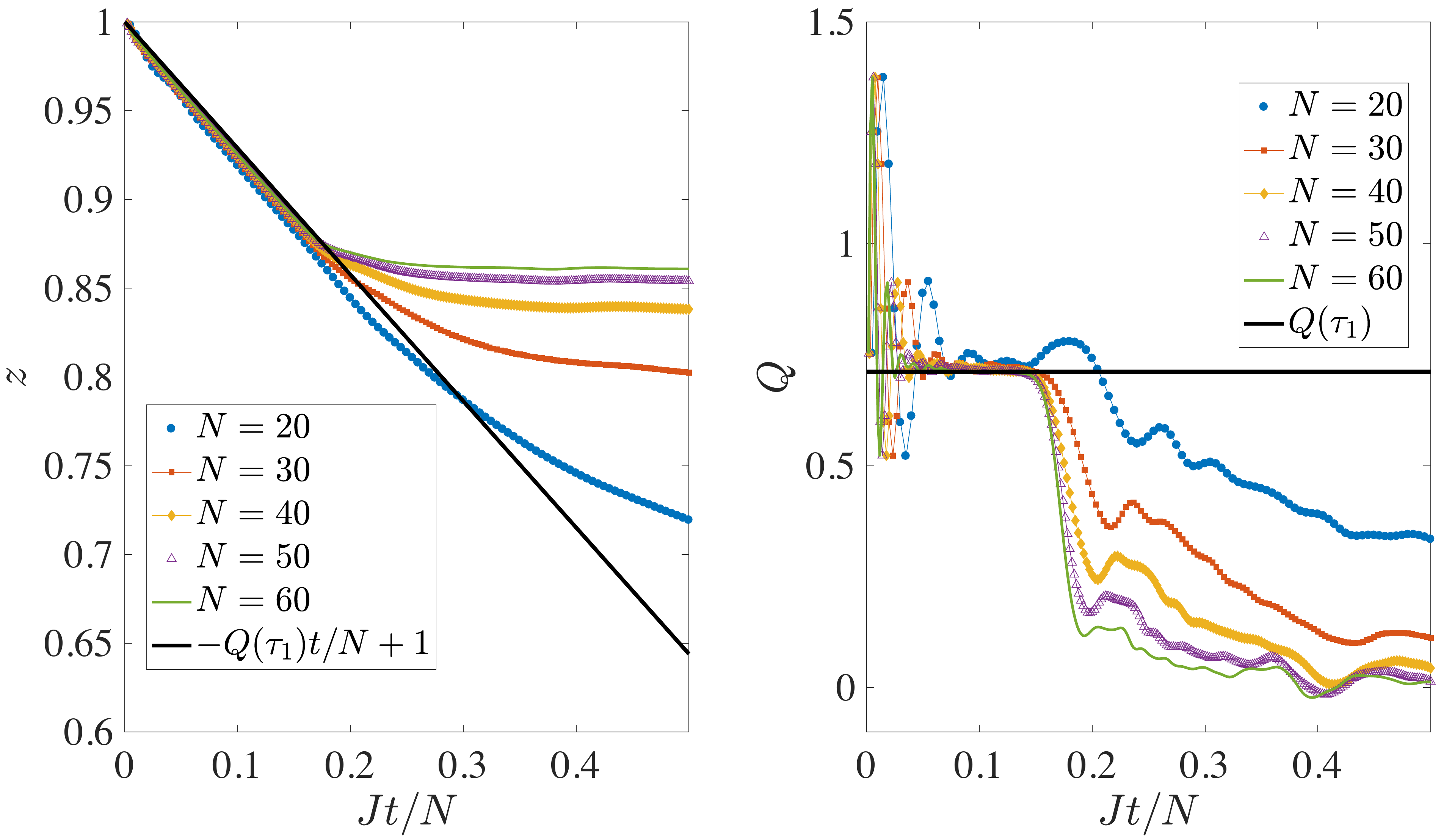}}\\ 
\caption{ (Left)~The battery magnetization as a function of time, as computed for different system sizes. (Right)~The current at the battery-system junction for $J_z=1.1J$.}
\label{ChainWBathsJz1p1}
\end{figure}

{There are two main standard approaches for characterising the type of transport: the spreading of perturbations~\cite{SJZP} and the system size scaling of steady-state currents~\cite{ZnidaricLowu,Prosen}. We note that both approaches may be directly linked as discussed in~\cite{PhysRevLett.117.040601} and in principle only one of the above criteria should be enough to characterize the transport. We have found in practice that the time behavior approach is more adequate for the strong coupling regime we address. 
The phenomenological model in Refs.~\cite{ZnidaricLowu,Prosen,PhysRevLett.117.040601} ensures non-zero steady state currents, even outside the ballistic regime. In contrast, our simulations explicitly model the bath and do not indicate that currents necessarily persist. Therefore, even though we make an effort to relate the current work to the finite size scaling in Ref.~\cite{Prosen} we find the time behavior of the current to be a more appropriate object for study.

Let us assume the current to scale with the system size as $Q\propto N^{-\alpha}$. If $\alpha=0$ the system is a perfect ballistic conductor, otherwise we have generic diffusion, with $\alpha<1$ indicating super-diffusion, $\alpha=1$ diffusion and $\alpha>1$ sub-diffusion.
Alternatively, in the spirit of spreading of inhomogeneities, it is interesting to study the demagnetization of the battery as a function of time. To this purpose, we study the up-magnetized lead, as the other side has symmetrically opposite dynamics. The total magnetization $z=\sum_i^{N_b}\langle Z_i\rangle/N_b$ of the lead can be expressed as a function of the spin current at the junction as
\begin{equation} z(t)=-\frac{1}{N_b}\int_0^{t}Q(\tau)d\tau+1.\end{equation} 
Let us assume for a moment that $Q(t)\propto t^{-\delta}$. In this case the standard criteria is such that
$\delta=0$ indicates ballistic transport, $\delta<\frac{1}{2}$ super-diffusion, $\delta=\frac{1}{2}$ diffusion and $\delta>\frac{1}{2}$ sub-diffusion.
}

The sudden connection of the battery to the system is a combination of two local quenches that spawn excitations. The dynamics induced by these quenches can be visualized by plotting the time dependent magnetization and currents. As shown in Fig.~\ref{ChainWBathsN60Jz0p5Jz1p1}, a double light-cone structure originates from the junctions, and is particularly manifest in the current profiles. In the ballistic regime, as soon as the two light cones meet they interfere constructively giving rise to an increased current. Following this time, the spatial profile of the magnetization along the system becomes flat. These features are typical signatures of ballistic transport. In contrast in the insulating regime, as soon as the second light cone reaches the opposite boundary, the spatial profile of the magnetization in the system displays staggered order superimposed to a domain-wall-like feature typical of the insulating regime~\cite{Prosen}. In remarkable contrast to the ballistic regime, the space-time diagrams show that the light-cone interference is destructive, leading to a decrease in the current.

Two very distinct dynamical regimes culminating at times $\tau_1$ and $\tau_2$ can be distinguished. The earliest regime is solely due to the dynamics generated by a single light cone, while the latest one arises from the interference between the two light cones. The earliest regime corresponds to constant currents at the battery-system junction such that $z(t)\approx-Q(\tau_1)t/N_b+1$. Here, $Q(\tau_1)$ displays a weak dependency on $J_z$ while remaining constant as the system size is increased. We therefore infer that $Q(\tau_1)$ represents the quasi-asymptotic current due to a single junction of two semi-infinite chains. As show in Fig.~\ref{ChainWBathsJz0p5}, for $J_z=0.5J$ a second regime sets in with the arrival of the second light cone at the junction, and a new constant current $Q(\tau_2)$ is established. This current is also independent of the system size indicating that the steady-state resulting from the two quenches is characterized by ballistic transport. In Fig.~\ref{ChainWBathsJz1p1} we show the data for $J_z=1.1J$. Once again, there is an initial ballistic regime in which the battery loses magnetization linearly in time. 
At the arrival of the second light cone, however, the loss of magnetization slows down and the resulting currents $Q(\tau_2)$ decreases significantly as the system size is increased.

\begin{figure}
{\includegraphics[width = 3.4in]{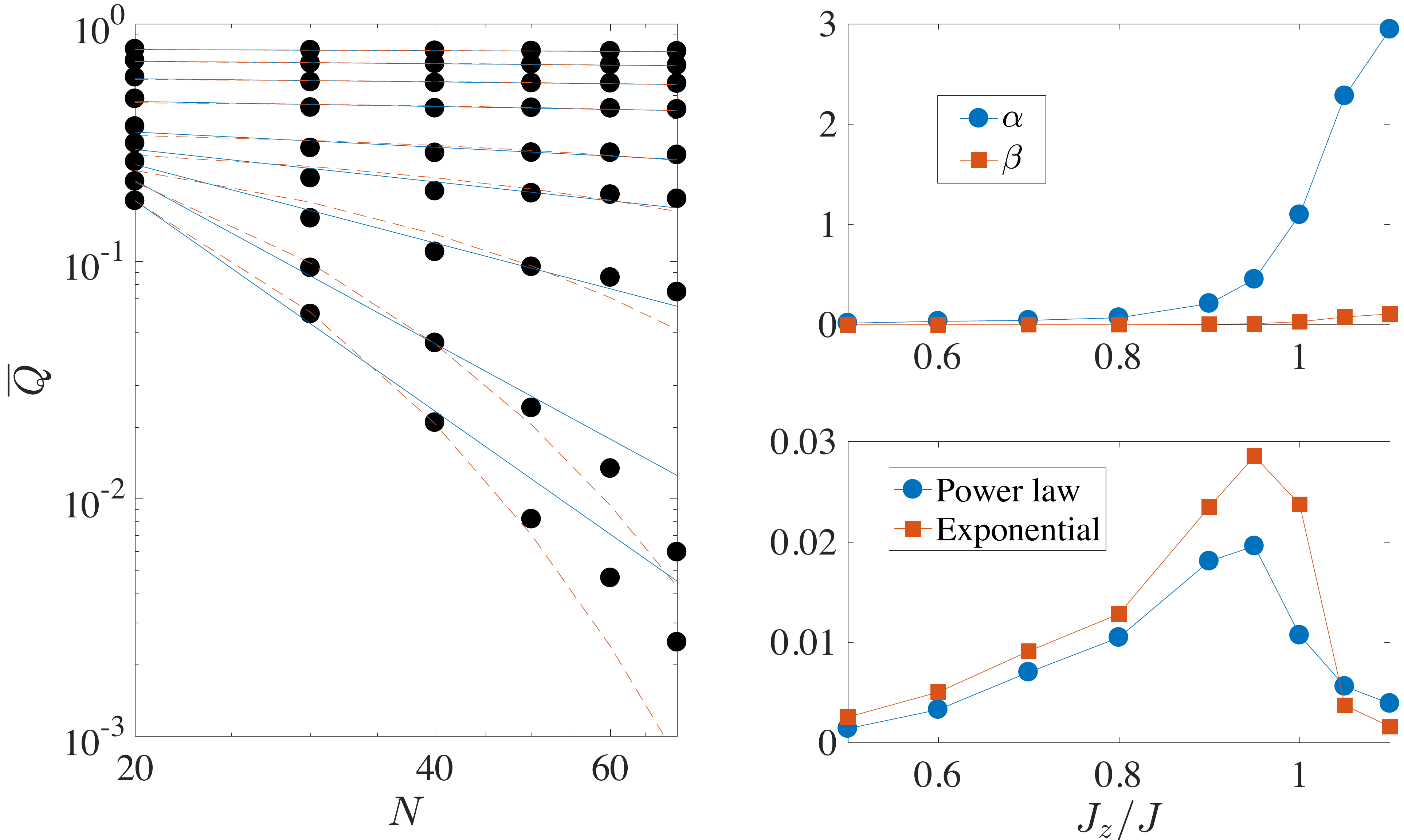}}\\ 
\caption{ (Left) Dots are data points while solid lines correspond to power law fits and dashed lines correspond to exponential fits. The interval for current averaging is taken as $[\tau_2=0.35, T=0.5]N/J$. (Upper right) The power law exponent $\alpha$ and the exponential rate $\beta$ obtained by fitting the current $Q_{\tau_2}(N)$ with $A N^{-\alpha}$ and $B e^{-\beta N}$ respectively. (Lower right) The corresponding errors in the fitting procedures.}
\label{ChainWBathsFits}
\end{figure}

The non-Markovian regime addressed here induces oscillations in the long time current around its average $\overline{Q}=\frac{1}{T-\tau_2}\int_{\tau_2}^TQ(t)dt$.
In order to characterize how the average current scales with system size, we have carried out fits both with a power law and an exponential. In Fig.~\ref{ChainWBathsFits} we show the exponents and the error associated to each fit. This allows us to compare and identify which scaling best approximates the raw data.
The vanishing exponents for $J_z<J$ suggest that the current converges towards a constant value in the limit $N\rightarrow\infty$, as is evident from from Fig.~\ref{ChainWBathsPhase}.
At the isotropic point we observe a super-diffusive exponent $\alpha>1$ and the error resulting from the power-law fit becomes significantly smaller then the one of the exponential fit. Above the Heisenberg point $J_z>J$ the current becomes small and susceptible to simulation errors, particularly to those induced by the truncation of the bond dimension. However, we observe that the exponential error becomes smaller than the power-law one. These features are indications of an insulating regime.
The conclusions drawn from this scaling analysis are summarized in the regime diagram of Fig.~\ref{ChainWBathsPhase}.

\begin{figure}
{\includegraphics[width = 3.4in]{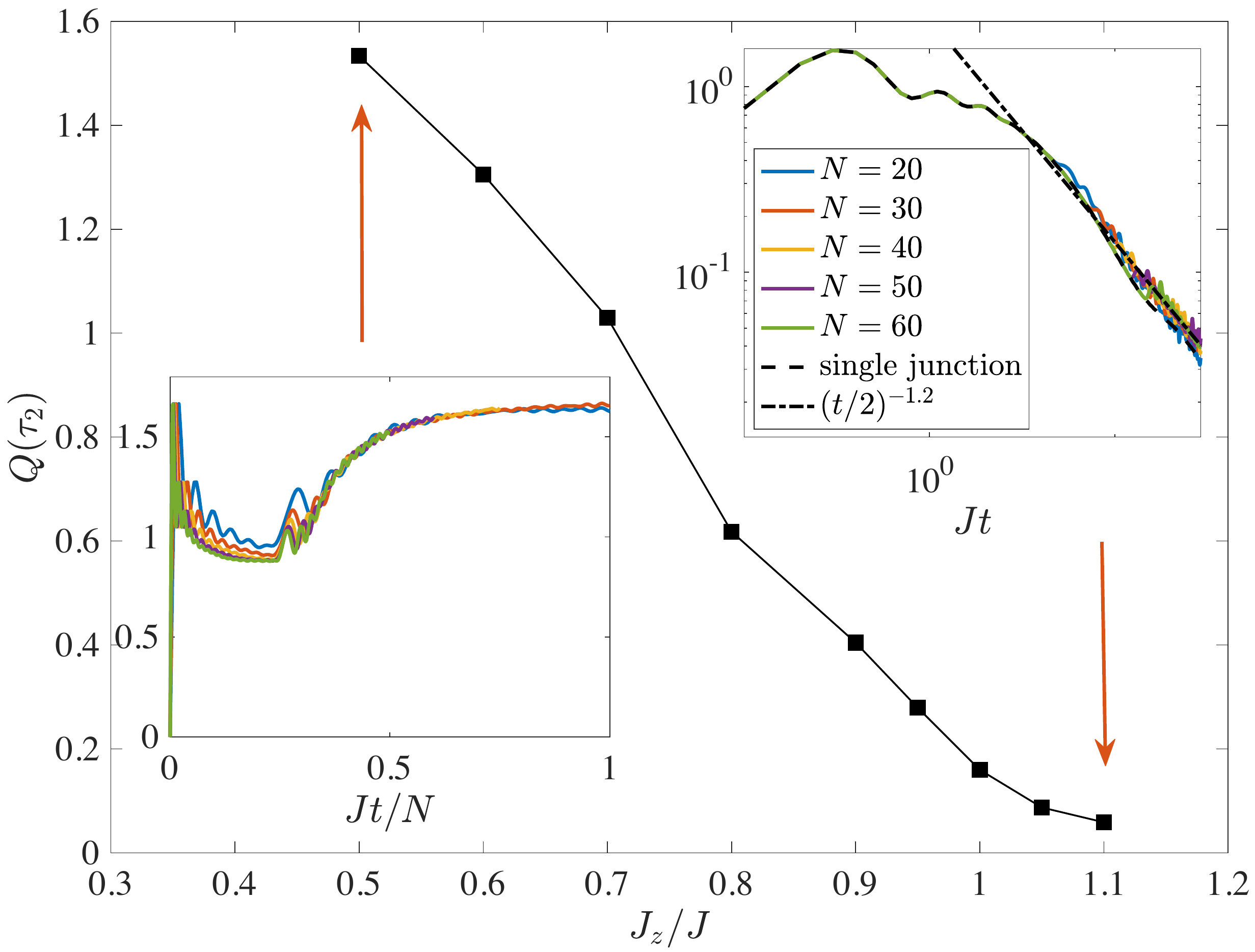}}\\ 
\caption{Currents $Q(\tau_2=0.5N/J)$ as a function of $J_z$ for the initial GHZ preparation. The system size is fixed to $N=40$, while each lead has $N_b=80$. (Inset)~Currents as a function of time for different system sizes, at two different values of $J_z$. Notice how for $J_z=1.1J$ the time has not been normalized by $N$, in order to compare with the dynamics of a single junction.}
\label{ChainWBathsPhaseGHZ}
\end{figure}

In order to investigate the choice of the initial state, we study the effect of preparing the system in the GHZ state, which has zero local magnetization but is orthogonal to the ground state. This state is at the other extreme of the spectrum residing in a high energy subspace. The long-time dynamics exhibits a different behavior, suggesting that even after long times the system bears memory of the initial state.  As shown in Fig.~\ref{ChainWBathsPhaseGHZ}, below the Heisenberg point the interference between the two light cones leads to an enhancement of current that is almost independent of system size. This is a similar behavior compared to the initial ground state preparation, suggesting that the system is in a ballistic regime. Instead, above the Heisenberg point, there is no destructive interference, and the dynamics at the interface follow very closely those of a single junction. 
A sub-diffusive signature is marked by the asymptotic current $Q(Jt\gg1)\approx(Jt/2)^{-1.2}$.
This suggest that the system goes through a weakly sub-diffusive regime before reaching the insulating steady state. In analogy with the data displayed in Fig.~\ref{ChainWBathsPhase}, the current as a function of the spin repulsion shown in Fig.~\ref{ChainWBathsPhaseGHZ} provides evidence that the regime diagram is similar for the two choices of the initial state, but that the dynamical properties exhibited are quite different.

{We would like to state that the size scaling done in Fig.~\ref{ChainWBathsFits} was done to provide the direct comparison between the Markovian regime addressed in [3,4] and the strong coupling limit of our work. The local phenomenological master equations adopted in [3,4] ensures non-zero steady state currents for finite $N$ even in the insulating regime, with the difference between regimes being how the non-zero steady state currents decay with system size. Our results in the strong coupling limit suggest that non-zero steady state currents might only remain in the ballistic regime. The main signature of the different regimes is then how the currents decay in time and not in system size, similarly to what has been done for the single junction of two chains. Our results indicate that the ballistic regime has non-zero steady state currents while other regimes present currents that vanish in time as shown in the inset of Fig.~\ref{ChainWBathsPhaseGHZ} and Fig.~\ref{ChainWBathsJz1p1} respectively. 
}

\begin{figure}
{\includegraphics[width = 3.4in]{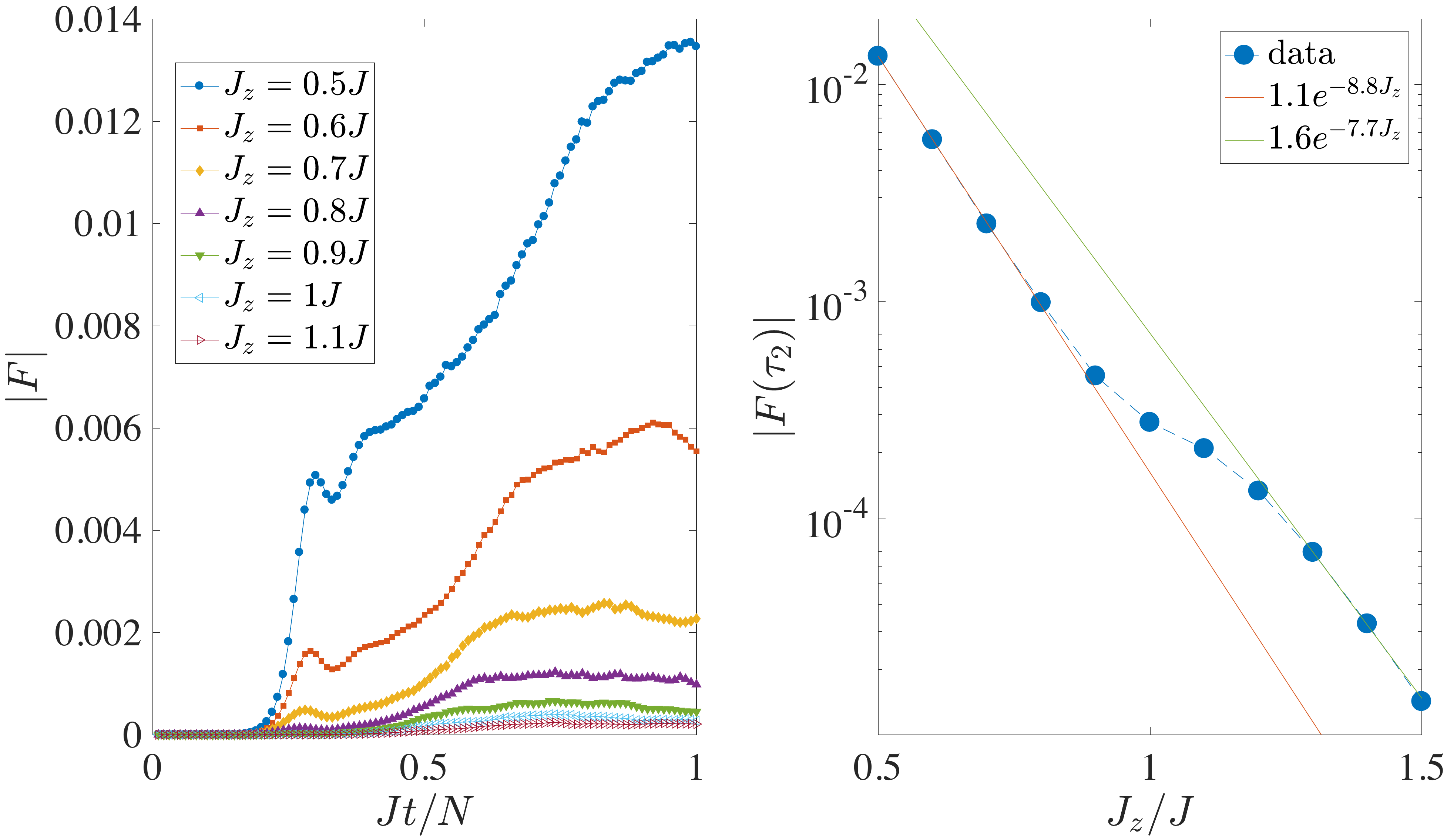}}\\ 
\caption{ (Left) The fidelity $F$ as a function of time, as computed for different values of $J_z$ and (Right) the corresponding $F(\tau_2)$ as a function of $J_Z$. Data taken for a system with $N_b=2N$ and $N=10$.}
\label{ChainWBathsMK}
\end{figure}

Inspired by quantum measures~\cite{BENM} of non-Markovian behavior, we perform a direct comparison between the two initial states by computing the ensuing time-dependent overlap. This can be done by first tracing out the bath degrees of freedom $\rho_S=\mathrm{tr}_B\{|\Psi_S\rangle\langle\Psi_S| \}$ and then computing the overlap as the fidelity $F=\langle \rho_{GHZ}|\rho_G\rangle/\sqrt{\langle \rho_{GHZ}|\rho_{GHZ}\rangle\langle \rho_{G}|\rho_G\rangle}$. The quantity $F$ is shown in Fig.~\ref{ChainWBathsMK}. Non-Markovian character is witnessed by the non-monotonic behaviour of $F$ as time increases. We further notice that in the ballistic regime the fidelity increases with time. In contrast, in the insulating regime the overlap between the two states remains small, due to the lack of transport. When characterized as a function of $J_z$, the long-time fidelity shows two distinct domains of exponential decay, respectively below and above the Heisenberg point, separated by a `sweet spot' at $J_z=J$. This behavior provides evidence of the different sensitivity of the two transport regimes to the initial state of the chain.

{Determining the precise universality class of the transition is a hard numerical task that escapes our current capabilities. Despite such limitations it is interesting to note that, if we contemplate the current at the junction as an order parameter, we would have that, in the thermodynamic limit, the ballistic regime has nonzero current while at and above the Heisenberg point the current is zero. From our results we cannot determine the universality class of the transition, we can only conjecture that it could be a continuous second-order-like transition, substantially departing from the ground state physics that has a BKT transition also at the Heisenberg point. Such a departure is however reasonable since the dynamics induced by the leads take the system far from the equilibrium ground state, allowing for genuine non-equilibrium phenomena.
}

In summary, we have presented a numerical study of the non-Markovian boundary-driven spin chain and characterized a nonequilibrium ballistic-diffusive-insulating regime transition. Our results provide clear evidence of the three distinct regimes in the thermodynamic limit. While the global features of the regime diagram bear close similarity to the case with Markovian leads, our analysis shows that the non-Markovian strong coupling between the system and the leads produces unique microscopic features with no analog in the open case. 
The setting considered in this work could be realized in practice with ultra-cold atoms in optical lattices, thanks to the remarkable progress in quantum state preparation and highly sensitive local measurements. A natural evolution of our study consists in characterizing the bias/interaction regime diagram, by assuming smaller magnetization differences between the two leads of the battery.
Lastly, it would also be interesting to characterize the interaction/disorder regime diagram in which we would expect a diffusive-insulating regime boundary even in the low bias regime.

\bibliographystyle{apsrev4-1} 
\bibliography{references}
\end{document}